\documentclass[]{epl}

\title{Scanning tunneling spectroscopy of a-axis \chem{YBa_2Cu_3O_{7-\delta}} films:
\emph{\textbf{k}}-selectivity and the shape of the superconductor
gap}

\shorttitle{Scanning tunneling spectroscopy of a-axis ...}

\author{A. Sharoni\inst{1} \and G. Leibovitch\inst{2} \and A. Kohen\inst{2} \and R. Beck\inst{2} \and G.
Deutscher\inst{2}\and \\ G. Koren\inst{3} \and O.
Millo\thanks{Corresponding author, e-mail:
\email{milode@vms.huji.ac.il}}\inst{1}}

\shortauthor{A. Sharoni \etal}

\institute{
  \inst{1} Racah Institute of Physics, The Hebrew University, Jerusalem 91904,
  Israel\\
  \inst{2} School of Physics and Astronomy, Raymond and Beverly Sackler Faculty
  of Exact Science, Tel Aviv University, 69978 Tel Aviv, Israel\\
  \inst{3} Department of Physics, Technion - Israel Institute
           of Technology, Haifa 32000, Israel
}

\pacs{74.50.+r}{Proximity effects, weak links, tunneling
 phenomena, and Josephson effects}
\pacs{74.72.Bk }{Y-based cuprates}
\pacs{74.80.-g} {Spatially inhomogeneous structures}

\begin{document}

\maketitle

\begin{abstract}
Tunneling spectroscopy of epitaxial (100) oriented
\chem{YBa_2Cu_3O_{7-\delta}} films was performed using an STM at
4.2 K.  On atomically smooth areas, tunneling spectra revealing
clear U-shaped gaps with relatively low zero bias conductance were
measured.  These spectra can be well fitted to the tunneling
theory into a d-wave superconductor only when introducing a strong
dependence of the tunneling probability on the wave-vector
\emph{\textbf{k}}. Possible origins for this
\emph{\textbf{k}}-selectivity in STM measurements will be
discussed.  On other areas, V-shaped gaps as well as zero bias
conductance peaks are observed, indicating relaxation of
\emph{\textbf{k}}-selectivity and the effect of nanofaceting,
respectively.
\end{abstract}

The \emph{d}-wave symmetry of the order parameter of
\chem{YBa_2Cu_3O_{7-\delta}} (YBCO) \cite{1,2,3} has a special
signature in the tunneling spectra measured for these
superconductors. The hallmark of this is the zero bias conductance
peak (ZBCP) that appears when tunneling along or near the nodal
direction ([110] for YBCO). This reflects the existence of Andreev
bound states at the Fermi level residing on the pair-breaking
(110) surface \cite{4,5,6,7,8}. The ZBCP was measured by numerous
groups, using both macroscopic and microscopic (STM) tunnel
junctions \cite{9,10,12,13,14,15,16}. Manifestation of the
\emph{d}-wave order parameter in the tunneling spectra acquired
along the anti-nodal ([100] for YBCO) and the c-axis directions is
more controversial. The theory of tunneling into a \emph{d}-wave
superconductor predicts a V-shaped gap structure for both
tunneling directions in the case where the density of states (DOS)
is isotropically averaged over the Fermi surface \cite{6,7}. Here,
quasi-particle excitations are observed at any finite energy,
since the nodal directions in \emph{\textbf{k}}-space are
monitored, resulting in a linear dependence of the dI/dV
\textit{vs.} V tunneling spectra on energy around the Fermi energy
(zero bias).  Such isotropic tunneling scenario is believed to
apply to STM measurements \cite{17}. In contrast, U-shaped gaps
are predicted to appear when \emph{\textbf{k}}-vectors along the
anti-nodal direction are preferentially monitored in the tunneling
process \cite{7,12,15,18,19}. Resolving this issue is thus
important for understanding STM spectroscopy in high temperature
superconductors and other anisotropic systems.

Experimentally, V-shaped gaps are indeed commonly observed in
c-axis tunneling measurements \cite{7,12,15,20,21,22,23}.
Recently, however, U-shaped gaps were observed in tunneling
spectra acquired on the (001) \chem{Bi_2Sr_2CaCu_2O_{8+\delta}}
(BSCCO) surface \cite{18}. This was accounted for by a
\textquoteleft \emph{\textbf{k}}-selection
mechanism\textquoteright ~inducing preferential tunneling along
the anti-nodal directions, resulting from the large overlap, and
consequently strong coupling, between the tip's electronic states
and the 3\emph{d} orbitals of the Cu atoms, through the mediation
of their 4\emph{s} states \cite{24}. However, a corresponding
effect was not yet established for the \emph{a}-axis surface,
where direct coupling to the Cu atoms is possible.

The data reported for a-axis tunneling is much more diverse,
possibly due to  problems in preparing well-defined (100) cuprate
surfaces, and keeping the exposed Cu-O planes from deterioration.
A wide range of features were observed, including zero bias peak
structures, mostly inside gaps \cite{12,14,25}, as well as
V-shaped \cite{26,27} and U-shaped gaps \cite{19,28}. The zero
bias conductance peak may result from nano-faceting or roughness
of the measured surface, as has been demonstrated by various
groups \cite{15,22,29,30}. The U-shaped gaps were accounted for by
assuming a narrow tunneling cone centered along the anti-nodal
direction, although, as will be shown below, the \textquoteleft
orbital coupling\textquoteright ~model discussed above may apply
as well.

In this paper we present an extensive scanning tunneling
spectroscopy study of (100) YBCO surfaces, directed at resolving
some of the questions arising from the contradictory experimental
findings described above.  We have measured films prepared using
two different deposition techniques, DC sputtering and laser
ablation. For both film types, spectra exhibiting U-shaped gaps
with relatively low zero bias conductance are found on smooth
areas. These spectra are well fitted to the tunneling theory into
a \emph{d}-wave superconductor only when introducing a tunneling
probability that is preferentially strong along the anti-nodal
directions. However, we cannot distinguish between the
\textquoteleft orbital coupling\textquoteright ~and the
\textquoteleft tunneling cone\textquoteright ~models. In areas
showing a rougher topography, only V-shaped gaps and zero bias
conductance peaks (ZBCP) of varying strength are observed.

\begin{figure}
\onefigure [width=14.5cm,height=3.6cm]{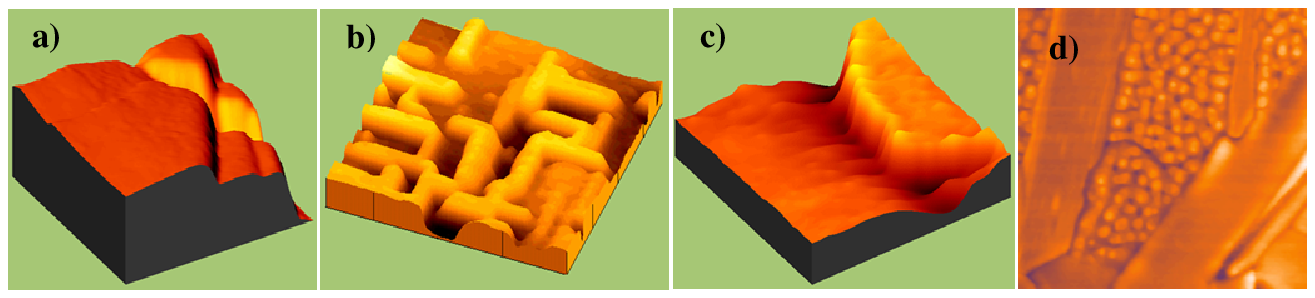} \caption{STM and
AFM images portraying the surface morphology  of YBCO films
containing various degrees of c-axis (background) and a-axis
(outgrowth) orientations. (a) and (b) - DC sputtered YBCO films
with about 70\% a-axis orientation, where (a) is a $180\times180
\un{nm^{2}}$ STM topographic image focusing on a single a-axis
crystallite, $25 \un{nm}$ in height,
 and (b) is a $1.6\times1.6
\un{\mu m^{2}}$ AFM image showing an array of outgrowing
crystallites, having atomically smooth areas on the scale of a few
tens of nm. (c) and (d) - laser ablated films, where (c) is a
$35\times35 \un{nm^{2}}$ STM image displaying a single outgrowing
a-axis crystallite, $20 \un{nm}$ in height, and  (d) is a
$2\times2 \un{\mu m^{2}}$ AFM image of a 95\% a-axis film
exhibiting two types of areas. The first is composed of small
crystallites a few unit cells in height, while the second consists
of large areas, atomically smooth on scales of $100 \un{nm}$. }
\label{fig1}
\end{figure}

Epitaxial thin YBCO films of various types were prepared by either
DC sputtering or laser ablation, with relatively broad
transitions, onset at 86 (88) K and ending at 82 (81) K for the
DC-sputtered (laser ablated) films. Both preparation methods
produced films yielding very similar spectroscopic data in spite
of the different morphology. \chem{YBa_2Cu_3O_{7-\delta}} with 5\%
Ca doping were grown on \chem{LaAlO_3} substrates by DC off-axis
sputtering, applying the method described in Ref. \cite{31} for
c-axis films. Here, however, the deposition temperature was
reduced to $720 ^{\circ}C$ in order to promote the development of
a-axis crystallite outgrowths \cite{32}.  These films consisted of
dense arrays of rectangular a-axis crystallites, typically 20 nm
in height and a few hundreds of nm in size, covering 70-90\% of
the surface area, as confirmed by X-ray diffraction, Atomic Force
Microscopy (AFM) and Scanning Electron Microscopy (SEM).  Many of
these crystallites had large atomically smooth areas on the top.
The STM and AFM images in Figs. \ref{fig1}a and \ref{fig1}b
exhibit the surface morphology of these films, showing a single
crystal (and a junction with its neighbor) and an array structure,
respectively. We have also measured DC sputtered a-axis YBCO films
prepared using the procedure described in Ref. \cite{25}, also
showing large flat areas.

The laser-ablated films were grown on (100) \chem{SrTiO_3} wafers
by two different dual-step processes.  First, a $30 \un{nm}$ thick
YBCO layer was prepared at a wafer temperature of $600 ^{\circ}C$
(or $650 ^{\circ}C$). Then, a second $75 \un{nm}$ thick YBCO layer
was deposited at $750 ^{\circ}C$ (or $785 ^{\circ}C$).  The films
prepared at the higher temperatures displayed a-axis crystallite
outgrowths on a c-axis background, similar to the DC sputtered
films, as depicted in the STM image of Fig. \ref{fig1}c. The lower
deposition temperature produced films with two coexisting a-axis
phases on about 95\% of the film area (as determined by X-ray
diffraction). One phase consisted of small crystallites, a few
unit-cells in height, while the other is composed of large areas,
atomically smooth on scales of hundreds of nanometers, as depicted
by the AFM image in Fig. \ref{fig1}d.

The samples were transferred directly from the growth chamber into
a chamber that was over-pressured by dry oxygen, then mounted
within a few hours into a home made cryogenic STM and cooled down
to 4.2 K in He exchange gas, being exposed to air for less than 15
minutes.  A total of 5 sputtered films and 4 laser-ablated films
were measured.  The tunneling dI/dV \textit{vs.} V curves were
acquired either directly by the use of conventional lock-in
technique, or by numerical differentiation of the measured I-V
curves, with similar results obtained by both methods. We have
confirmed that the measured gaps and ZBCP features were
independent of the STM voltage and current setting (before
disconnecting momentarily the feedback circuit).  This rules out
the possibility that the gap features are due to the Coulomb
blockade, which is known to be sensitive to these parameters
\cite{33}. All the STM data presented in this paper were acquired
at 4.2 K with (normal) tunneling resistances between $100
\un{M\Omega}$ to $1 \un{G\Omega}$.

\begin{figure}
\onefigure [width=12cm,height=5cm]{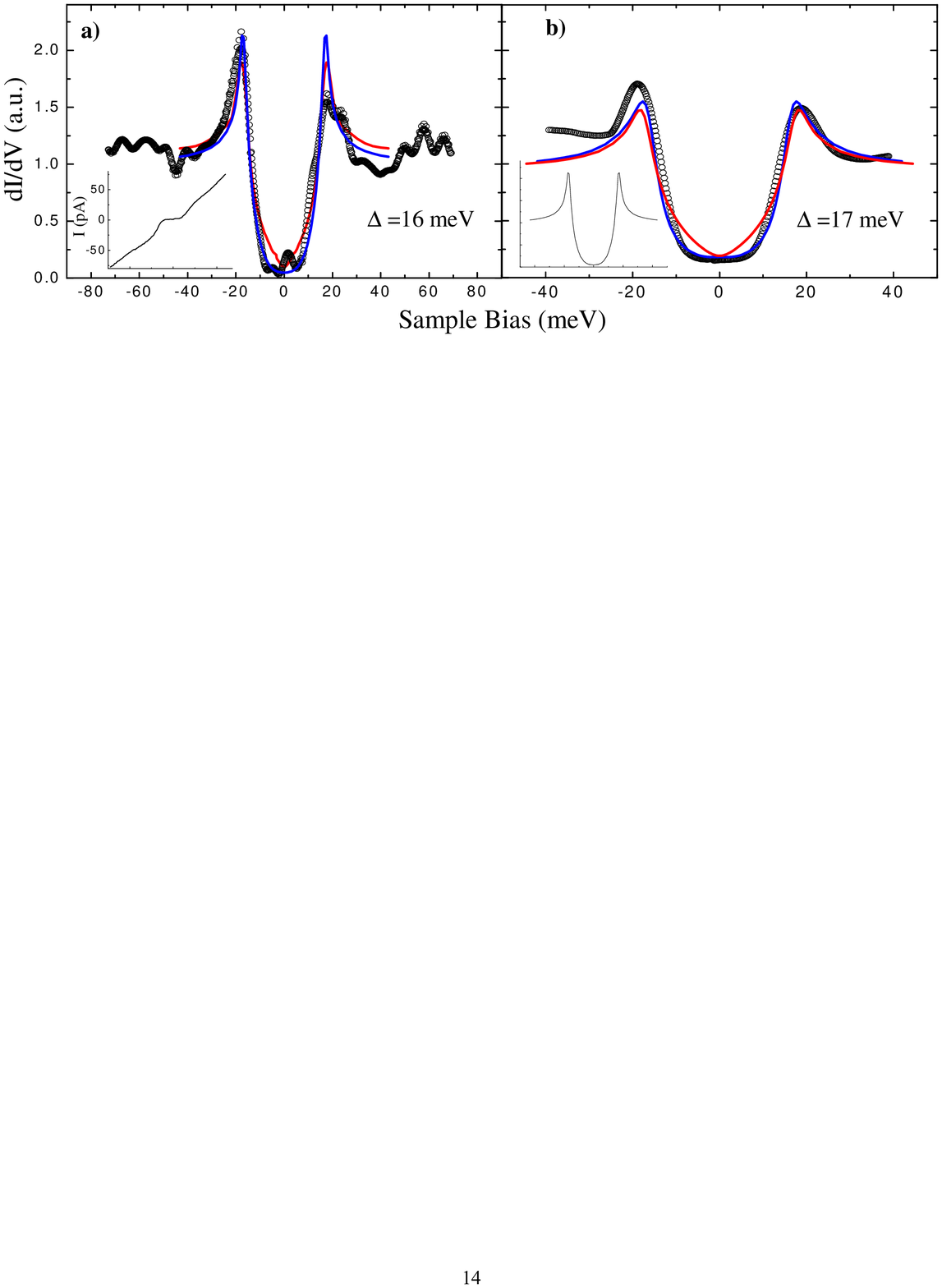} \caption{a) A
tunneling spectrum measured on the a-axis crystallite shown fig.
\ref{fig1}c (empty circles), with the corresponding I-V curve
shown in the inset. The blue curve is a fit using the
\textquoteleft orbital coupling\textquoteright ~model with a $16
\un{meV}$ gap and $\Gamma=0.06\Delta$. b) A spectrum measured on
the crystallite shown in fig. \ref{fig1}a (empty circles), fitted
to the \textquoteleft narrow tunneling cone \textquoteright
~model, $\Delta=17 \un{meV}$ and $\Gamma=0.1\Delta$. An equally
good fit was obtained for the \textquoteleft orbital
coupling\textquoteright ~model, shown in the inset. For
comparison, we present spectra calculated assuming isotropic
averaging over the Fermi surface, with the same $\Delta$ and
$\Gamma$ values, showing a large deviation from the measured
spectra at the low bias region (red curves).} \label{fig2}
\end{figure}

On all the measured samples we have found atomically smooth areas,
on the scale of tens of nanometers or more, where the dI/dV
\textit{vs.} V curves featured mainly U-shaped gaps, with
relatively low zero bias conductance. The gap value distribution
in these regions was relatively narrow, varying between 16 to 18
meV, for both laser ablated and DC sputtered films. Typical
measurements are presented in fig. \ref{fig2} (empty circles),
where spectrum \ref{fig2}a was measured on the laser-ablated film,
on the a-axis crystallite shown in fig. \ref{fig1}c, and spectrum
\ref{fig2}b is a representative example for the sputtered films,
acquired on the crystallite shown in image \ref{fig1}a.  The inset
of fig. \ref{fig2}a presents the I(V) curve from which the
differential conductance was derived, depicting a BCS-like
structure.  The red curves were calculated using the theory for
tunneling into \emph{d}-wave superconductors \cite{6,7}, taking an
equal weight for all \emph{\textbf{k}}-vectors, i.e., an isotropic
tunneling matrix. It is clear that the V-shaped gap structure thus
obtained deviates considerably from the experimental curves at the
low energy regime, the experimental curves being by far more flat.
The blue curves in fig. \ref{fig2}, on the other hand, were
calculated assuming preferential tunneling in the anti-nodal
directions.  In fig. \ref{fig2}a we used the \textquoteleft
orbital coupling\textquoteright ~model, whereas in fig.
\ref{fig2}b the \textquoteleft narrow tunneling
cone\textquoteright ~approach was applied, both reproducing well
the measured U-shaped gap structure.

For the \textquoteleft orbital coupling\textquoteright ~model we
used the tunneling matrix element given by Misra et al. \cite{18},
$|M(k)|^{2} \propto [\cos(k_xa)-\cos(k_ya)]$, where $k_x$ and
$k_y$ are the wave
vector components along the main axes in the
a-b plane. Note that this model was applied in Ref. \cite{18} to
c-axis tunneling. However, it should also be valid for tunneling
into the (100) surface, where the lobe of the Cu \emph{d}-orbital
is perpendicular to the surface, thus strongly overlapping
directly with the tip states.  The blue curve in fig. \ref{fig2}a
was calculated using this model, with a gap of $\Delta = 16
\un{meV}$ and a small Dynes broadening parameter \cite{34},
$\Gamma = 0.06\Delta$, showing good agreement with the
experimental data in the gap region. Equally good fits were
obtained using the \textquoteleft narrow tunneling
cone\textquoteright, described below.

The \textquoteleft narrow tunneling cone\textquoteright ~model is,
in a sense, a semi-classical approach. Within the WKB
approximation, and at bias voltages much smaller than the
tunneling barrier height $\Phi$, the tunneling transition
probability decays exponentially with increasing $k_T$, the
transverse component of the wave vector \emph{\textbf{k}}:
$p(\emph{\textbf{k}}) \propto exp[-\{\hbar^{2}d/(2m\Phi)
^{1/2}\}k_T^{2}]$, and consequently $p(\theta) \propto
exp[-\beta\sin^{2}(\theta)]$. Here, $\theta$ is the angle between
the momentum of the tunneling electron and the surface normal and
$\beta = \sqrt{2m/\hbar^{2}}\frac{E_F}{\sqrt{\Phi}}d$ , where
$E_F$ is the Fermi energy and d is the width of the tunneling
barrier \cite{12,19,35,36}. For typical tunneling junctions, this
will result in a narrow tunneling-momentum cone around the surface
normal ($\Delta\theta \sim 20^{\circ}$, full angle).

This type of reasoning doesn't alter the characteristics of
tunneling in the [110] or the [001] direction, but predicts a
U-shaped gap for the (100) surface.  Figure \ref{fig2}b
demonstrates a fit obtained using this model with $\Delta=17
\un{meV}$ and $\Gamma= 0.1\Delta$. This particular fit was
calculated with $\Delta\theta = 20^{\circ}$, but no significant
difference is found in the range of $0^{\circ} < \Delta\theta <
30^{\circ}$. In the inset we present, for comparison, a nearly
identical spectrum calculated using the \textquoteleft orbital
coupling\textquoteright ~model. We were not able, in this work, to
differentiate between these two models, but it is clear that a
\emph{\textbf{k}}-selection mechanism is essential to account for
our results.

It should be pointed out that V-shaped curves obtained from the
isotropic tunneling model, such as those presented in red in fig.
\ref{fig2}, can be smeared using very large Dynes broadening
parameter until a structure resembling U-shaped gaps would emerge.
Then, however, the zero bias conductance becomes much larger than
that of the experimental data.

\begin{figure}
\onefigure [scale=0.55]{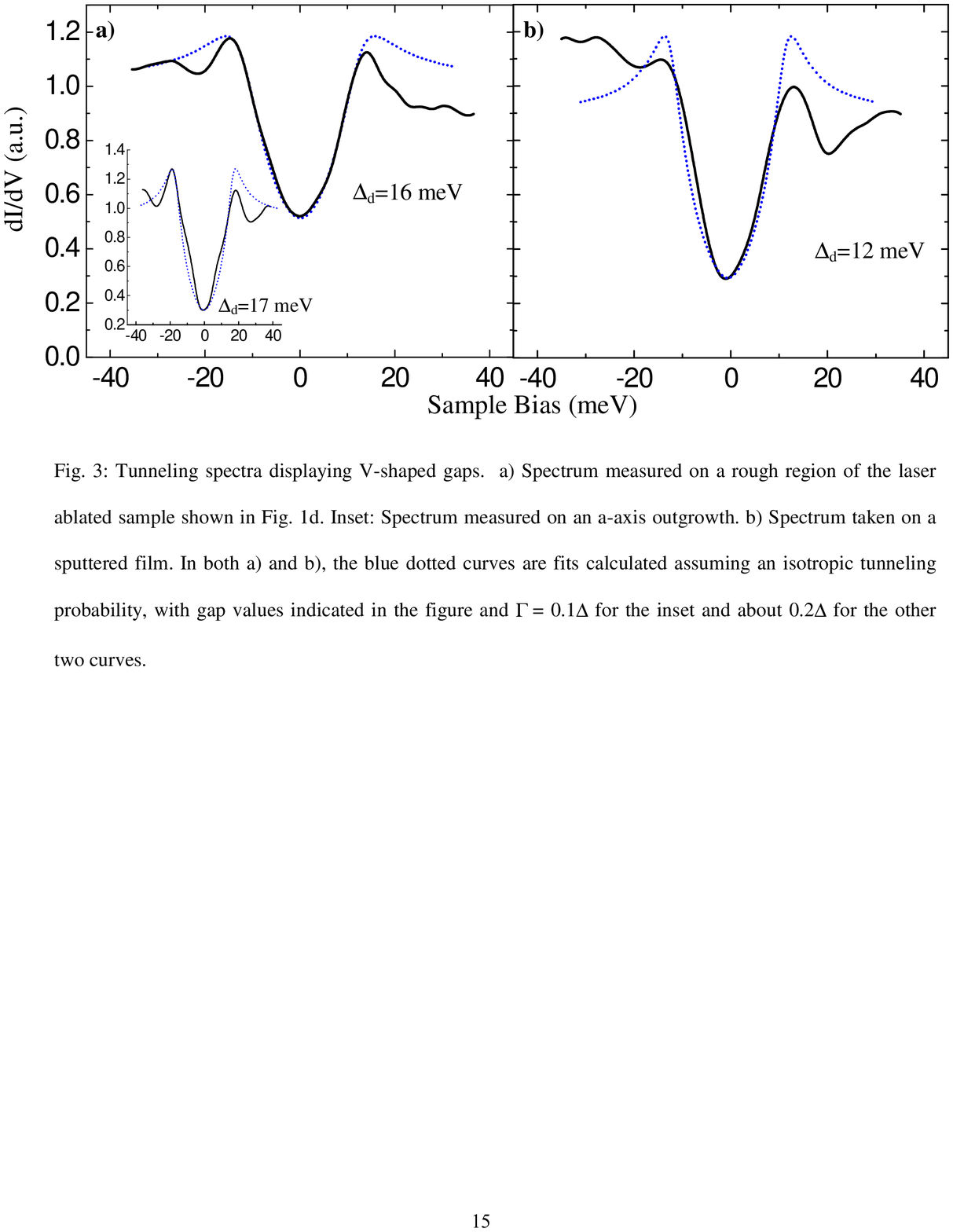} \caption{Tunneling spectra
displaying V-shaped gaps.  a) Spectrum measured on a rough region
of the laser ablated sample shown in fig. \ref{fig1}d. Inset:
Spectrum measured on an a-axis outgrowth. b) Spectrum taken on a
sputtered film. In both a) and b), the dotted curves are fits
calculated assuming an isotropic tunneling probability, with gap
values indicated in the figure and $\Gamma = 0.1\Delta$ for the
inset and about $0.2\Delta$ for the other two curves.}
\label{fig3}
\end{figure}

In areas having a larger degree of surface roughness and disorder,
U-shaped gaps were never observed, and the spectra showed only
V-shaped gaps or ZBCPs. The zero bias conductance in these curves
were typically larger as compared to the U-shaped spectra, varying
between $20\%$ to $50\%$ of the normal differential conductance
(at high bias). Best fits of these data are obtained using the
isotropic tunneling-probability model with a lifetime broadening
of around $0.2\Delta$. Fits with any of the
\emph{\textbf{k}}-selective models described above, yield
unreasonably large Dynes parameters of more than $0.5\Delta$.  In
fig. \ref{fig3} we present typical V-shaped gap structures
measured on various film types, along with the corresponding fits
depicted by the dotted curves, with gap values denoted in the
figure.  The spectra in fig. \ref{fig3}a were measured on the two
types of laser-ablated films. The one in the mainframe was taken
on a rough part of the film shown in fig. \ref{fig1}d, and the
other in the inset is for an a-axis crystallite outgrowth, both
showing a fairly low zero bias conductance. The data in
\ref{fig3}b was acquired on an a-outgrowth from a DC sputtered
film. The relatively small gap in this figure may be due to local
deviations from optimal doping resulting from losses of oxygen
from the surface. The gap distribution in these regions were much
larger as compared to the smooth ones, varying between 12 to 17
meV, with no apparent dependence on film preparation method.

As mentioned earlier, the U-shaped gaps were found only on smooth
areas, whereas V-shaped structures are typically measured on more
disordered regions.  A question then arises regarding the role
played by the local surface morphology on the measured spectra.
Since the typical gap size for both gap shapes (and also in the
presence of a ZBCP, see below) is the same, around $17 \un{meV}$,
it appears that the local surface roughness does not affect much
the order parameter at the surface. We note in passing that our
data are in agreement with Ref. \cite{37}, where similar gap
values were obtained for both [110] and [100] orientations,
further confirming the conclusion \cite{37} that YBCO is in the
\emph{d}-wave weak coupling regime. However, surface disorder is
known to relax conservation rules for the in-plane component of
the \emph{\textbf{k}}-vector, and in our case, may wash out the
strong \emph{\textbf{k}}-dependent tunneling probabilities
introduced in either the \textquoteleft orbital
coupling\textquoteright ~or \textquoteleft narrow tunneling
cone\textquoteright ~models. Surface roughness seems also to
affect the quasi-particle life-time at the surface, smearing the
gap structure and thus enhancing the Dynes parameters needed for
the fits.

\begin{figure}
\onefigure [scale=0.46]{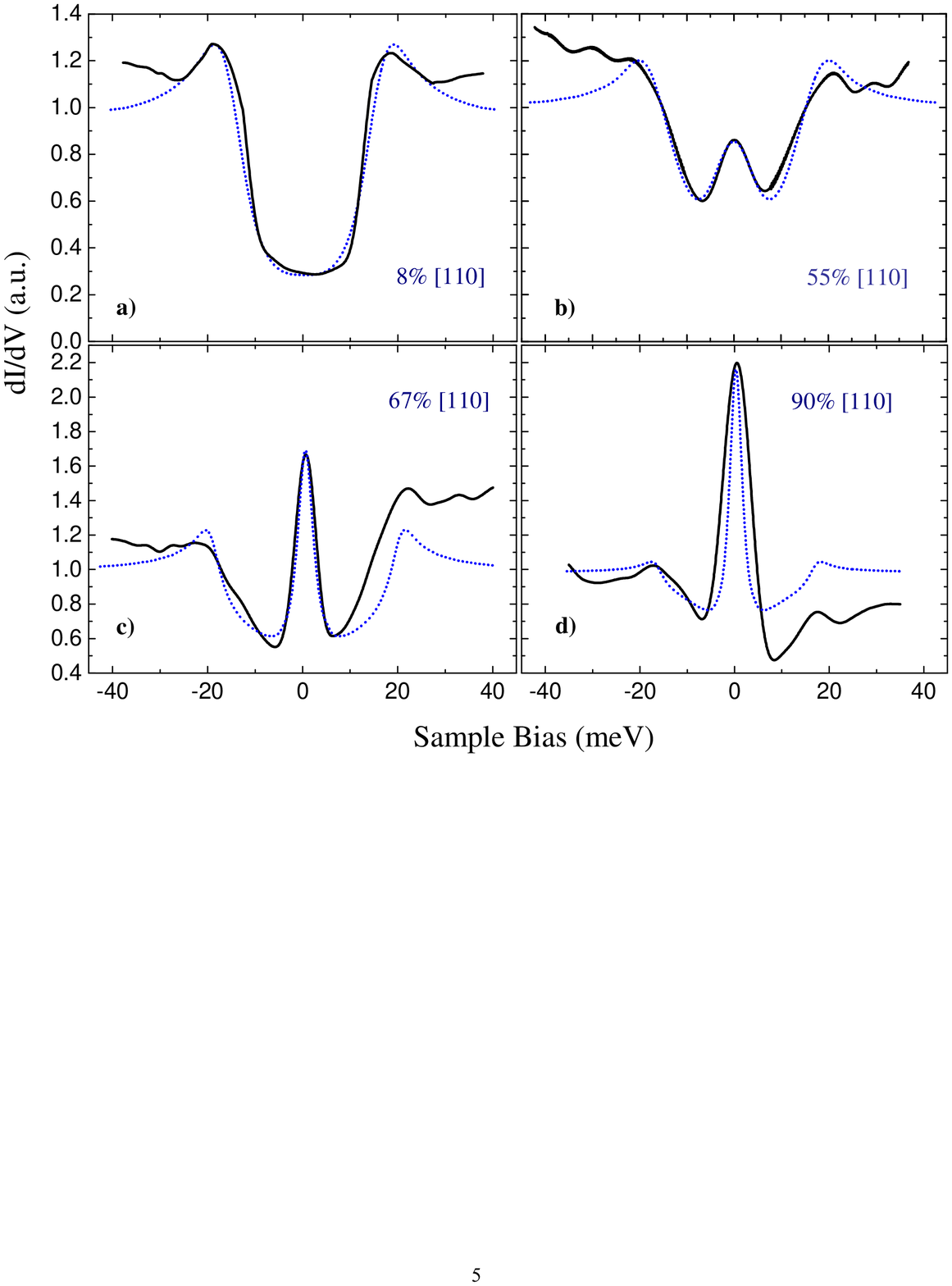} \caption{dI/dV \textit{vs.} V
curves measured on a faceted region on the YBCO film shown in fig.
\ref{fig1}d. The curves can be fitted well by a \textquoteleft two
channel\textquoteright ~tunneling model, assuming independent
contributions of tunneling in the [110] and [100] directions. The
same $\Delta=17\un{mV}$ and $\Gamma=0.15\Delta$ were used in all 4
panels, and the different relative contribution of [110] tunneling
is indicated in each panel.  The fits presented here were
calculated with the \textquoteleft narrow tunneling
cone\textquoteright ~model, but similar fits are achieved with the
\textquoteleft orbital coupling\textquoteright ~model.}
\label{fig4}
\end{figure}

Another typical spectral feature that we have found is a ZBCP
embedded in a gap structure, resembling dI/dV-V characteristics
obtained for macroscopic tunnel junctions on (100) YBCO surfaces
\cite{14,25}. This behavior was attributed to the effect of (110)
facets. Indeed, STM measurements have shown that even a facet of
unit-cell height can give rise to a ZBCP \cite{15,29}. Such a
local sensitivity was not yet demonstrated for the (100) YBCO
surface. In fig. \ref{fig4} we present spectra exhibiting
\textquoteleft peak in a gap\textquoteright ~structures measured
on a faceted (rough) region of the laser ablated film shown in
fig. \ref{fig1}d, where the relative ZBCP height increases from
\ref{fig4}a (no visible peak) to \ref{fig4}d. The curves can be
fit well by employing a \textquoteleft two channel\textquoteright
~model, where independent contributions of tunneling in the [110]
and [100] directions are taken into account.  The magnitude of the
[110] tunneling component varies spatially, as is indicated in
each figure, whereas the gap and the Dynes parameters were nearly
the same, $\sim 17 \un{meV}$ and $0.15\Delta$, respectively. The
fit to the U-shaped gap in fig. \ref{fig4}a requires a
\emph{\textbf{k}}-selective tunneling process. However, due to the
relatively large zero bias conductance level (as compared to other
U-shaped gaps), a small [110] contribution had to be included;
(the peak structure is washed out due the broadening parameter).

In conclusion, our data provide clear evidence that the local
tunneling spectra acquired for a-axis YBCO films do not always
reflect the bulk superconductor DOS averaged isotropically over
the Fermi surface.  Tunneling characteristics exhibiting U-shaped
gap structures, rather than the V-shapes, are often measured. The
observation of U-shaped gaps does not depend on the sample
preparation method, but is correlated with the degree of surface
smoothness. Two possible models that reproduce equally well these
spectra were considered: a narrow tunneling cone around the [100]
normal direction, and a strong overlap of the tip electronic
states with the \emph{d}-wave electronic orbitals of the Cu ions
emerging out of the surface, both enhance tunneling preferentially
along the anti-nodal directions.

\acknowledgments This work was supported in parts by the Israel
Science Foundation, the Heinrich Hertz-Minerva Center for High
Temperature Superconductivity, and by the Oren Family Chair for
Experimental Solid State Physics.

\end{document}